\documentclass{PoS}

\newcommand{\swf}{{\it Swift}}

\newcommand{\xrt}{{\it XRT}}
\newcommand{\uvt}{{\it UVOT}}
\newcommand{\bat}{{\it BAT}}

\newcommand{\latf}{Fermi-{\it LAT}}

\newcommand{\whipple}{{\it Whipple}}

\newcommand{\mrk}{Mrk~421}

\usepackage{aas_macro} 
\usepackage{natbib}
\usepackage{graphicx}
\usepackage{subfigure}
\usepackage{txfonts}
\usepackage[pdftex]{graphicx}
\usepackage{epstopdf}

\title{The Large Flares of Mrk 421 in 2006: 
signature of electron acceleration and energetic budget of the jet}

\ShortTitle{The Large Flares of Mrk 421 in 2006}

\author{\speaker{Andrea Tramacere}
        ISDC, Data Centre for Astrophysics Chemin d'Ecogia 16 CH-1290 Versoix Switzerland\\
        E-mail: \email{andrea.tramacere@unige.ch}}

\FullConference{The Extreme sky: Sampling the Universe above 10 keV\\
		 October 13-17 2009\\
		 Otranto (Lecce) Italy}
	
\abstract{
We present the results of a deep spectral analysis of all 
Swift observations of Mrk 421 between  April 2006 and July 2006, 
when it reached its highest X-ray flux recorded until the end of 2006. 
We completed this data set with other historical X-ray observations.
We used the full data set to investigate the correlation between the 
spectral parameters.\\
We found a  signature of stochastic acceleration in the anticorrelation 
between the peak energy ($E_p$) of the spectral energy distribution (SED) and 
the spectral curvature parameter ($b$). We found signature of energetic 
budget of the jet in the correlation between the peak flux of the SED ($S_p$) 
and  $E_p$. Moreover, using simultaneous \swf~ UVOT/XRT/BAT data, 
we demonstrated, that during the strongest flares, the UV-to-X-ray 
emission from \mrk~ requires that the curved  electron distribution 
develops a low energy power-law  tail.\\
The observed spectral curvature and its anticorrelation with $E_p$ is 
consistent with both stochastic acceleration or energy-dependent 
acceleration probability mechanisms, whereas the power-law slope of 
\xrt-\uvt~ data is  close to that inferred from the GRBs X-ray 
afterglow and in agreement with the \textit{universal}  
first-order relativistic shock acceleration models. This scenario implies 
that magnetic turbulence may play a twofold role: spatial diffusion 
relevant to the first order process and momentum diffusion relevant to 
the second order   process.\\
 }
\FullConference{The Extreme sky: Sampling the Universe above 10 keV - extremesky2009,\\
		October 13-17, 2009\\
		Otranto (Lecce) Italy}

\begin{document}

\begin{figure}[t]
\begin{center}
\includegraphics[angle=0,width=9cm]{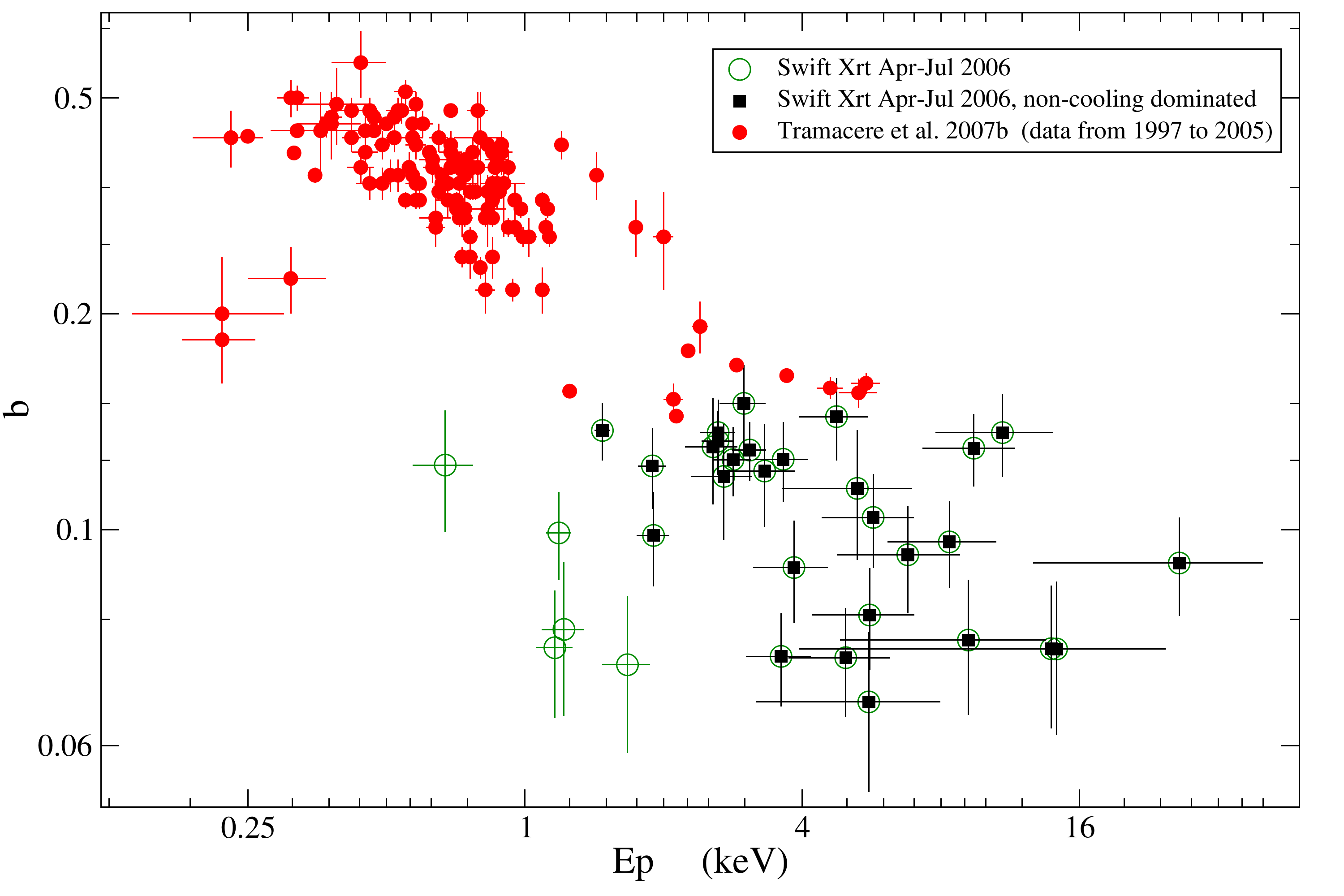}
\end{center}
\caption{Scatter plot of the curvature ($b$) vs. $E_p$. 
Solid red circles represent data from \cite{Trama2007b}. Black boxes represent Swift data from  
\cite{Tramacere2009}, without the cooling-dominated events. Empty circles 
represent the whole XRT data set presented in \cite{Tramacere2009}.}
\label{fig:Epb}
\end{figure}

\section{Introduction}
BL  Lac objects are Active Galactic  Nuclei (AGNs)  characterized 
by a polarised and highly variable nonthermal continuum emission 
extending from radio to  $\gamma$-rays. In the most accepted scenario, 
this radiation is produced within a relativistic  jet that originates 
in the central engine and points close to our line  of sight. 
The   relativistic outflow has a typical bulk Lorentz factor of 
$\Gamma \approx 10$, hence the emitted fluxes, observed at an angle 
$\theta$,  are affected by a beaming factor $
\delta = 1/(\Gamma  (1 - \beta \cos\theta ))$. \\
The Spectral  Energy Distribution  (SED) of these objects  has a typical 
two-bump  shape.  According to current models, the  lower-frequency 
bump is interpreted as synchrotron (S) emission from highly relativistic  
electrons with Lorentz factors $\gamma$ in excess of $10^2$. This 
component peaks at frequencies ranging from the IR to the X-ray band.
In the framework of the Synchrotron Self Compton (SSC) emission 
mechanism,   the  higher-frequency  bump can be attributed to 
inverse Compton scattering of synchrotron photons by the same 
population of relativistic electrons that produce the  synchrotron 
emission \citep{Jones1974}.\\ 
\mrk~ is classified as a High energy peaked BL Lac (HBL) \citep{Padovani1995} 
because its synchrotron emission peak ranges  from a fraction of 
a keV to several keV. With its redshift $z$ = 0.031, it is among the 
closest and most well studied HBLs. 
In spring/summer 2006,  \mrk~ reached its highest X-ray flux recorded until 
that time. The peak flux was about 85 milli-Crab in the 2.0-10.0 keV band, and 
corresponded to a peak energy of the spectral energy distribution (SED) that 
was often  at energies higher than 10 keV.\\ 
In this paper we use \swf~ UV and X-ray data \citep{Tramacere2009} of the 2006 flaring activity,
completed with historical X-ray observations, to interpret
the correlation between the spectral parameters 
in terms of acceleration processes and energetic budget of the jet.
 We remand the reader to  \citet{Tramacere2009} that paper for a more complete picture.

\begin{figure}[t]
\begin{center}
\includegraphics[angle=0,width=9cm]{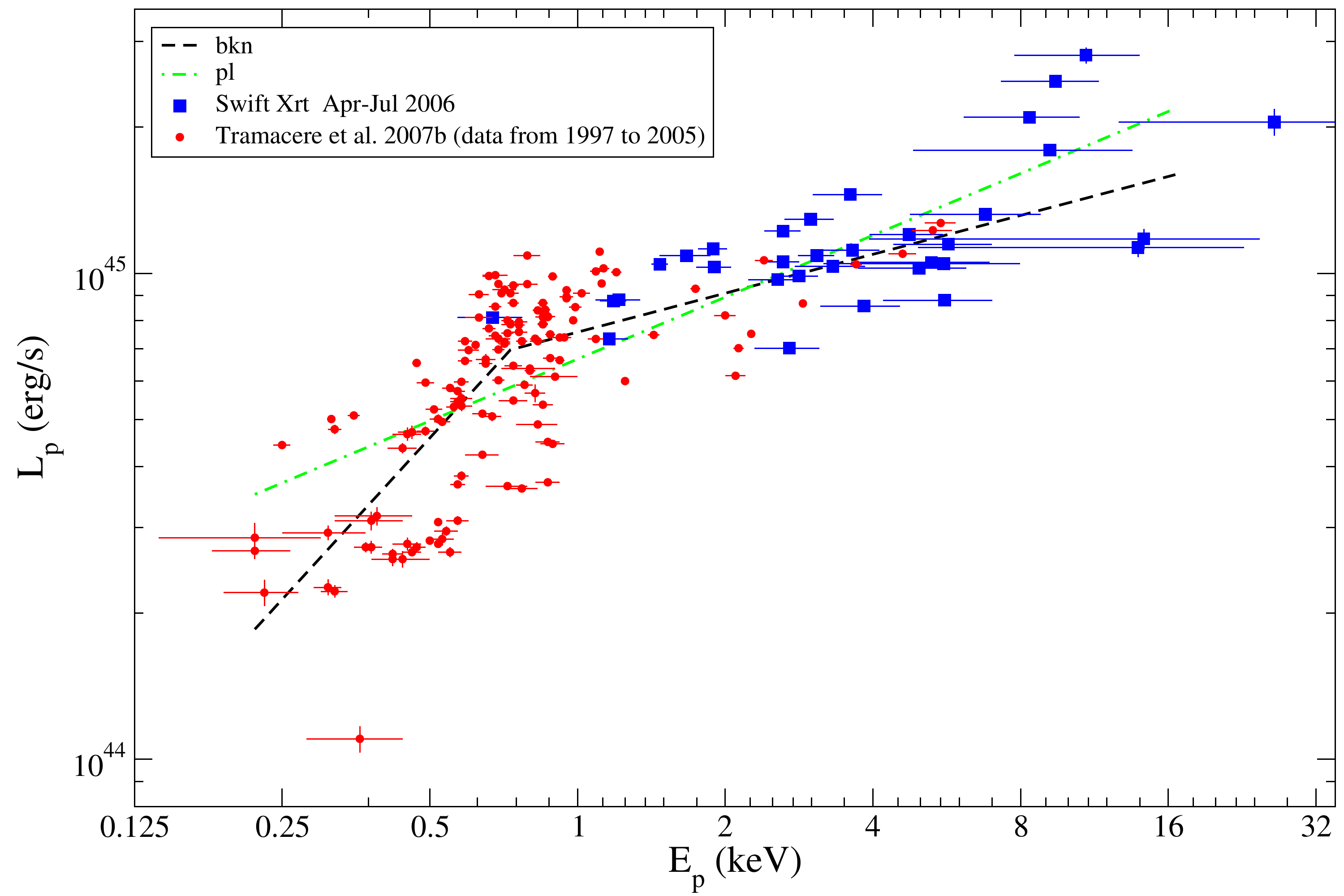}
\end{center}
\caption{Scatter plot of $L_p$ vs $E_p$.  
Solid red circles represent data from \cite{Trama2007b} spanning from 1997 to 2005. 
Solid blue boxes represent Swift data from \cite{Tramacere2009}. 
The green dashed dotted line represents the best fit using a power-law and the 
black dashed line represents the best fit using a broken power law.}
\label{fig:SpEp}
\end{figure}

\section{Spectral parameter trends}
\mrk~ showed in the past several major flaring episodes marked by 
flux variations that went along with significant spectral 
variations \citep{Massaro2004a}. The X-ray spectral shape in general 
exhibited a marked curvature that is described well by a log-parabolic model \citep{Massaro2004a, Trama2007b},
that can be expressed in terms of the SED peak energy ($E_p$), of the SED peak flux ($S_p$),	
and of the peak spectral curvature ($b$) as:
$S(E)= S_p~10^{-b~(\log(E/E_p))^2}$.\\
In the following we investigate the signature of stochastic acceleration 
in the anticorrelation between $E_p$ and $b$. Moreover,  we look for signatures 
of energetic budget of the jet in the correlation between $S_p$ and $E_p$.

\subsection{$E_p-b$ trend}
In Fig. \ref{fig:Epb} we show the $E_p-b$ trend  
for both \xrt~ 2006, and historical data. Thanks to the very high statistics
exhibited during the 2006 flaring activity, we were able to perform 
an highly temporally-resolved spectral analysis and to identify 
cooling-dominated  observations (empty circles in Fig. \ref{fig:Epb}), 
characterized by a strong flux decrease and strong spectral softening.
These cooling-dominated states are disentangled from the full dataset
because they can bias the acceleration-driven trend.\\ 
The $E_p-b$ scatter plot shows a clear anticorrelation, 
with  the peak energy of the S component increasing as its spectral 
curvature is decreasing.  
This trend hints for an acceleration processes producing 
curved electron distributions, where the curvature decreases as the
acceleration becomes  more efficient. A first possible scenario 
is that in which the acceleration probability of the particle is
a decreasing function of its energy \citep{Massaro2004a}. 
An alternative explanation is provided by the stochastic acceleration (SA) framework \citep{Karda1962,Trama2007b}. 
In the SA mechanism the momentum-diffusion term ($D$) is responsible
for the broadening of the electron energy distribution ($n(\gamma)$), hence it is inversely
proportional to the curvature in the observed X-ray photons ($b$) and to the 
curvature in the distribution of the emitting particles ($r$) . 
According to the relation between $r$ and $b$ 
\citep[$r\sim5b$]{Massaro2004a} we derive the following relation  among 
the peak energy of the electron distribution ($\gamma_p$), $~b$, and $E_p$:
\begin{equation}
\log(E_p) =2 ~\log(\gamma_p)+ 3/(5b).
\label{eq:Ep_gammap}
\end{equation}
Clearly, this trend predicts the anticorrelation between $E_p$ and $b$, 
in agreement with the observational data reported in Fig. \ref{fig:Epb}.

\subsection{$S_p-E_p$ trend}
The $S_p-E_p$ trend  (Fig. \ref{fig:SpEp} ) 
demonstrates the connection between the average energy of the particle 
distribution and the power output of the source. To obtain a deeper
understanding of the jet energetics, we plot on the $y$ axis  
$L_p = S_p4\pi D_L^2$,  where  $D_L\simeq 134$ Mpc is the luminosity 
distance\footnote{We used a flat cosmology model with $H_0=0.71$ 
km/(s Mpc) $\Omega_M=0.27$ and $\Omega_{\Lambda}=0.73$.}. 
In the case of synchrotron emission, we expect:
$S_p \propto n(\gamma_{3p}^3) B^2 \delta^4$ and $E_p \propto \gamma_{3p}^2 B \delta$,
where $\gamma_{3p}$ is the peak of $n(\gamma)\gamma^3$, $B$  
is the magnetic field, and $\delta$ is the beaming factor.\\
It follows that the dependence of 
$S_p$ on $E_p$ can be expressed in the form of a power-law: 
$S_p \propto Ep^{\alpha}$. The simple power-law fit gives a value of 
$ \alpha= 0.42 \pm 0.06$, this value clearly rules out as main drivers 
both $B (\alpha=2)$ and $\delta (\alpha=4)$, indicating $\gamma_p$ as 
the main driver.\\ 
A more detailed analysis of the scatter plot
reported in Fig.\ref{fig:SpEp} shows that the trend has a break at 
about 1 keV. The broken power law fit gives two slopes of 
$\alpha_1 = 1.1\pm 0.2$ and $\alpha_2 = 0.27 \pm 0.07$, respectively. 
This break in the trend implies that for $E_p\lesssim 1$ keV and $L_p\lesssim10^{45}$ 
erg/s, the driver follows the relation with $\alpha\simeq 1.0$ (we define 
this state the quiescent sate), whilst for $E_p\gtrsim 1$ keV
and $L_p>10^{45}$ erg/s, the driver relates to $\alpha\simeq 0.2$ (we define 
this to be the high state).\\ 
A possible interpretation is that the break of $S_p-E_p$ depends on 
the modulation of the number of emitted particles ($N=\int n(\gamma) d\gamma$).
Since $B$ and $\delta$ have been ruled out as main 
drivers of the $S_p-E_p$ trend, we assume that they have a small
variance. In this scenario, as $\gamma_{3p}$ (namely $E_p$) is increasing  
$N$ is constant up to a maximum value of the electron energy density
$u_e=\int \gamma m_ec^2 n(\gamma) d\gamma$ corresponding to a $L_p$
of  $\simeq 10^{45}$ erg/s, that could represent the maximum 
energetic budget of the jet. Above the value of $E_p\simeq$ 1 keV, $N$  decreases 
to don't exceed the maximum energy content.

\begin{figure}[t]
\begin{center}
\begin{tabular}{lr}
\includegraphics[angle=0,width=6cm]{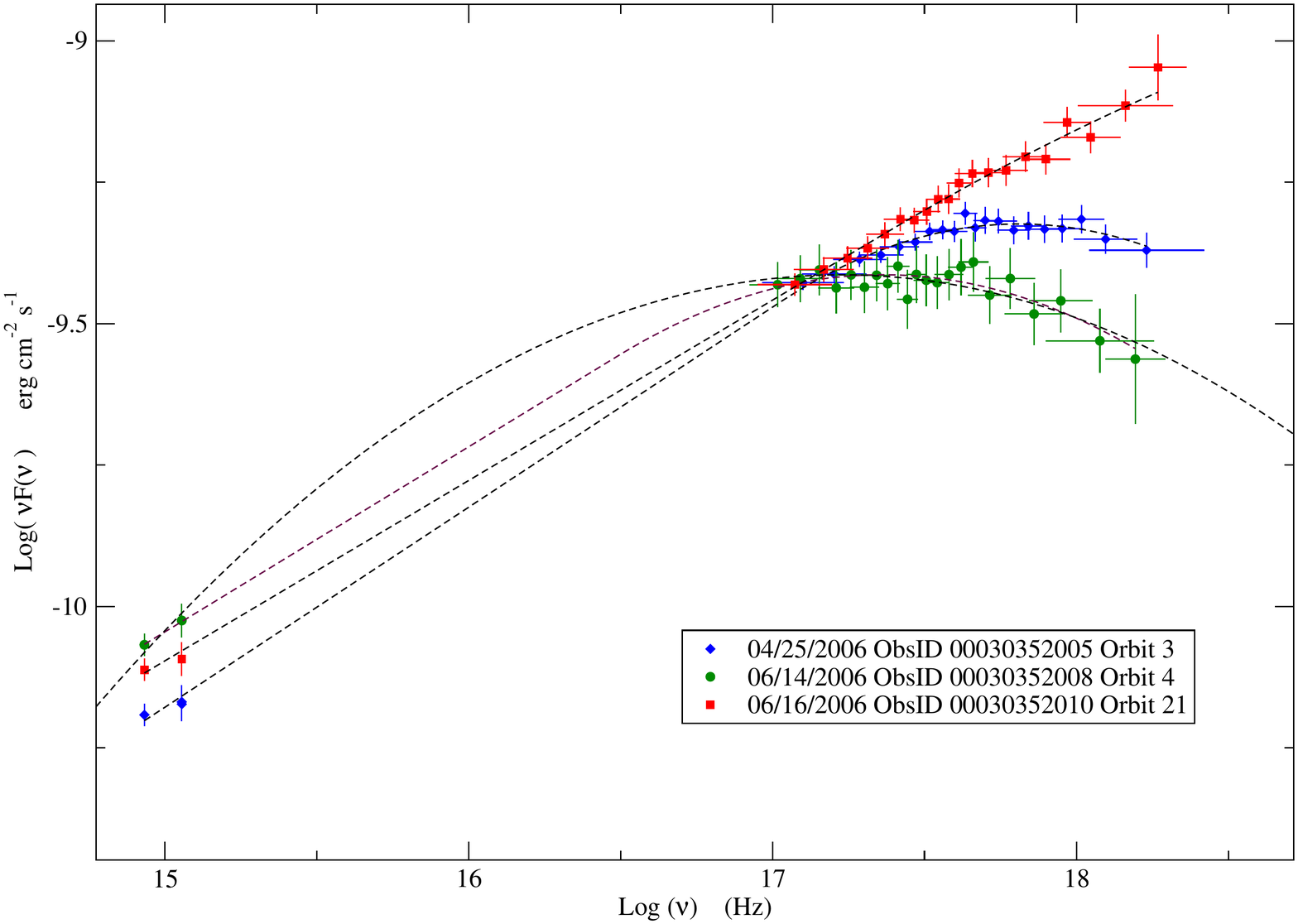} &
\includegraphics[angle=0,width=6cm]{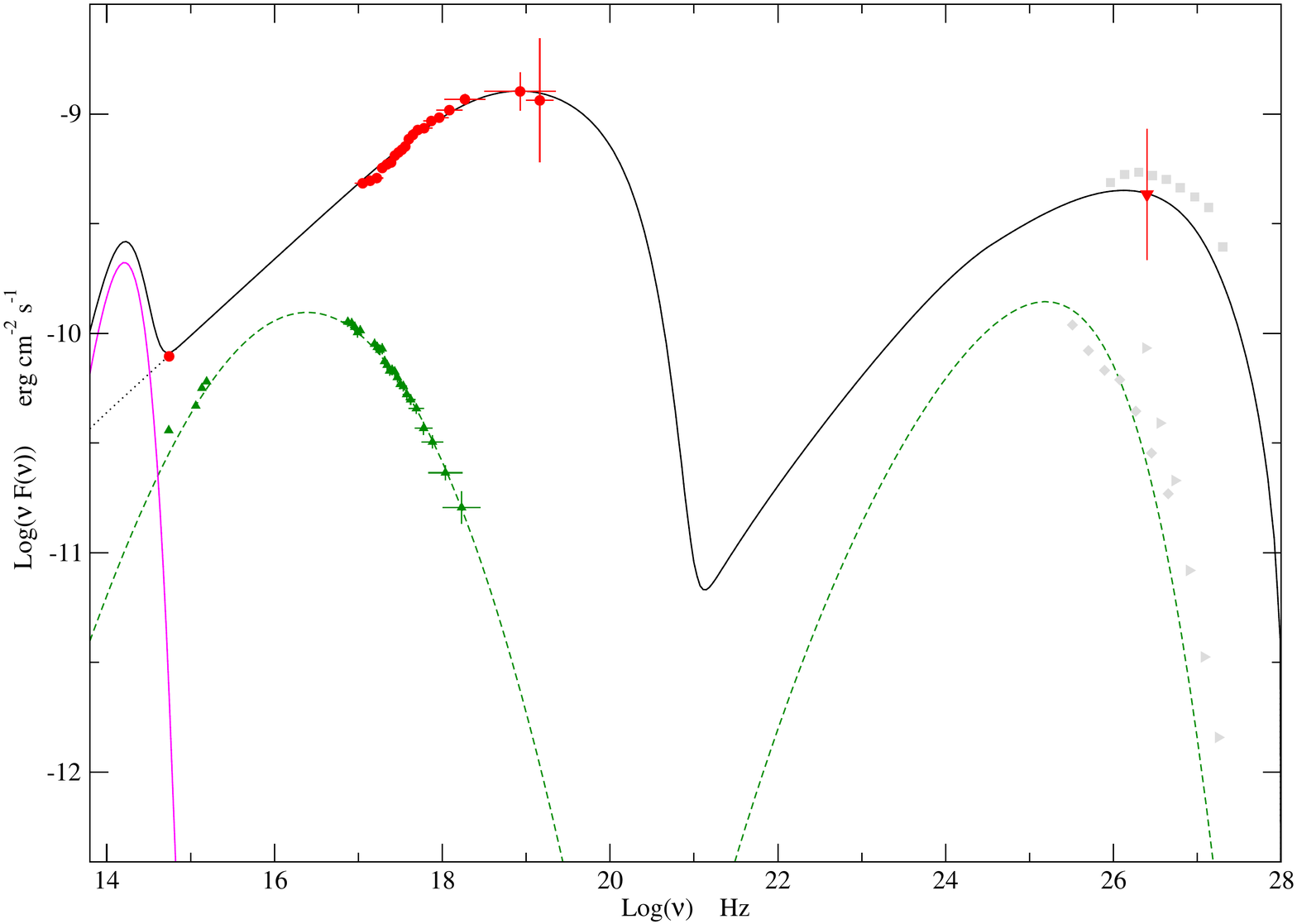} \\
\end{tabular}
\end{center}
\caption{\textit{Left} panel: Three different spectral shapes of the
UV-to-X-ray data from \cite{Tramacere2009}. Red boxes represent a PL 
spectrum  observed on 2006 June 16. Blue diamonds represent a LPPL 
spectrum on  2006 April 25. Green circles represent a LP spectrum  on 
2006 June 14. \textit{Right} Panel: SSC model. Solid red circles represent 
data from, 06/23/2006  \uvt-\xrt-\bat~ observations. Green up triangles 
shows \uvt-\xrt~ data on 03/31/2005 from \cite{Trama2007b}. Empty grey 
polygons represent non-simultaneous EBL corrected TeV data 
\citep{Albert2007,Yadav2007}. The solid red down triangle represent 
a \whipple~ observation on 18,19 and 21 June 2006 \citep{Lich2008}.
Best fit parameters for the 06/23/2006 state: source size 
$R=2.1\times 10^{15}$ cm, $B$=0.1 G, $\delta=25.0$,	 N=15.0 cm$^{-3}$, 
$s$=2.3, $~r$=0.65, $\gamma_c$=2.85$\times 10^5$.}
\label{fig:SED-UVOT-XRT}
\end{figure}

\section{UV-to-X-ray spectrum and SED modelling }
Combined simultaneous \uvt~ and \xrt~ observations 
\citep{Tramacere2009} shows   SEDs (see left panel of 
Fig. \ref{fig:SED-UVOT-XRT}	) that can be 
classified in three categories: \textit{a)} described by a 
log-parabola, \textit{b)} described by a power law,
\textit{c)} described  by a spectral law that is a power law at 
its low energy tail, becoming  a log-parabola function at its high 
energy one (LPPL)\\ 
The power-law spectral index,  in the $\nu F(\nu)$ representation, for model
\textit{b} and \textit{c}  were found in the range $a_{\nu}\simeq[0.25-0.4]$. 
According to the standard synchrotron theory, these values are
not compatible with a high value of the minimum energy of the electron 
distribution ($ \gamma_{min} $). Indeed, in this case we would 
expect $a_{\nu}\simeq 4/3$, a value much harder than the observed 
one. This implies a low energy power-law tail in $n(\gamma)$. 
Using the well known relation \citep{Ryb1979} between  the electron
energy distribution spectral index ($s$ ) and that in 
the S photons ($s=3-2a_{\nu}$) we obtain 
$s\simeq[2.2-2.5]$. A phenomenological description 
of $n(\gamma)$, consistent with the observed UV-to-X-ray
spectrum,  is given by the following spectral law:
\begin{eqnarray}
\label{eq:lppl-elec}
 n(\gamma)&=&K~(\gamma/\gamma_c)^{-s},~~~~~~~~~~~~~~~~~~ \gamma \leq \gamma_c \nonumber  \\
 n(\gamma)&=&K~(\gamma/\gamma_c)^{-(s+r~\log(\gamma/\gamma_c))} , \gamma >\gamma_c~~ ,
\end{eqnarray}
where $r$ is the curvature and $\gamma_c$  the turn-over energy.
The presence of  a power-law feature and the range of observed  
spectral indices  are relevant both in the context of Fermi 
first-order acceleration models and from an observational point of view.
Indeed,  our range of spectral index values ($ s \simeq [2.2-2.4]$ ) is 
consistent with the predictions  of the relativistic-shock acceleration ($s\simeq 2.3$), 
both in the case of analytical or numerical  approaches \citep{Achterberg2001,Blasi2005}.
Moreover, we note that  \citet{Waxman1997},  studying the  the afterglow X-ray emission of $\gamma-$ray 
bursts (GRB), inferred an electron distribution index of  $s\simeq 2.3 \pm 0.1$.  
Although the emission scenarios assumed in the case of GRBs are more complex
w.r.t. the one-zone homogeneous SSC model, the similar values of the electron energy 
spectral index hints for the  relativistic-shock acceleration as an underlying acceleration 
mechanism common both to GRBs and HBLs.
We remind that the power-law feature  can also be consistent with a purely stochastic scenario,
but in this case, it requires a fine tuning of the ratio of the acceleration 
to the loss timescale ($s\simeq 1+t_{acc}/t_{esc}$) to get a \textit{universal} index.\\  
An example of SSC modelling obtained using a LPPL electron distribution
is reported in the right panel of Fig. \ref{fig:SED-UVOT-XRT} 
(for a more detailed analysis of the SED modelling we remand to \cite{Tramacere2009}). 
The red points represent the simultaneous \swf~ data on 06/23/2006, and 
the green points data for the 31/05/2005 state. 
Here we just stress the huge difference in the spectral shape and flux levels
between the two states. Indeed, for the 2005 data the S component is well 
described both in the UV and X-ray window, by using as $n(\gamma)$ a pure 
log-parabola. On the contrary, the 2006 state requires a LPPL electron 
distribution to fit the UV-to-X-ray data. Because of the dominant galaxy 
contribution below  UV energies, 2005 \swf~ data don't allow to understand 
whether or not a low energy PL feature develops. We remark that such a 
feature could be tested comparing the prediction from a one zone homogeneous 
SSC model with simultaneous \swf~ and \latf~ MeV/GeV data.


\section{Conclusion and Discussion}
We have shown that the X-ray flaring activity of Mrk 421 results in a complex
spectral evolutions due to drastic changes in the electron energy
distribution probably related to a complex acceleration scenario.
Indeed, in our analysis we found both signatures of first
and second order acceleration processes acting at the same time.
The $E_p-b$ trend is consistent with a SA scenario with the X-ray spectral 
curvature related to the acceleration rather than to the cooling process.
The presence of a power-law low-energy tail, found during the brightest 
X-ray flares in 2006, and the corresponding values of the electron distribution 
index ($s\simeq[2.2-2.4]$) are consistent with the predictions of relativistic 
Fermi first-order acceleration models ($s\simeq 2.3$).
Our findings hint for a simultaneous role of the first 
and second order processes both related to the magnetic field
turbulence. The stochastic acceleration hence is related to a 
 momentum diffusion coefficient  
which drives  the curvature consistently with the $E_p-b$ observed
trend. 
The first order acceleration, observationally signed
by the low-energy power-law spectral index, is linked to the spatial 
diffusion coefficient.
Interestingly, recent  particle-in-cell simulations by \cite{Spit2008}
obtain electron distributions that are compatible with this scenario.\\ 
The $S_p-E_p$ trend demonstrates the connection between the average energy
of the particle distribution and the power output of the source.
The observed values of the expected power-law dependence 
($S_p \propto E_p^{\alpha}$) exclude $  B$, and $ \delta $, 
indicating $ \gamma_p $ as the main driver of the $S_p-E_p$ trend. 
The break in  the $S_p-E_p$ scatter plot (see Fig. \ref{fig:SpEp}) at about 1 keV, 
where the typical source luminosity is about $L_p \simeq 10^{45}  $ erg/s, 
can be interpreted as an indicator  of the energetic content of the jet.

\end{document}